\documentclass[boxit]{JAC2003}

\usepackage{graphicx}
\usepackage{booktabs}
\usepackage{verbatim}
\usepackage{subcaption}
\begin{document}

\title{SPP Beamline Design and Beam Dynamics}

\author{G. Turemen\thanks{gorkem.turemen@ankara.edu.tr}, Z. Sali, B. Yasatekin, Ankara University, Ankara, Turkey\\ 
V. Yildiz, Bogazici University, Istanbul, Turkey\\ 
M. Celik, Gazi University, Ankara, Turkey\\ 
A. Alacakir, TAEK-SANAEM, Ankara, Turkey\\ 
G. Unel, UCI, Irvine, California, USA\\
O. Mete, UMAN, Manchester, UK}

\maketitle

\begin{abstract}
  The Radio Frequency Quadrupole of SANAEM Project Prometheus will be a demonstration and educational machine which will accelerate protons from 20 keV to 1.5 MeV. The project is funded by Turkish Atomic Energy Authority and it will be located at Saraykoy Nuclear Research and Training Center in Ankara. The SPP beamline consists of a multi-cusp $H^+$ ion source, a Low Energy Beam Transport line and a four-vane RFQ operating at 352.2 MHz. The design studies for the multi-cusp ion source (RF or DC) were performed with IBSimu and SIMION software packages. The source has already been produced and currently undergoes extensive testing. There is also a preliminary design for the solenoid based LEBT, POISSON and PATH were used in parallel for the preliminary design. Two solenoid magnets are produced following this design. The RFQ design was made using LIDOS.RFQ.Designer and it was crosschecked with a home-grown software package, DEMIRCI. The initial beam dynamics studies have been performed with both LIDOS and TOUTATIS. This paper discusses the design of the SPP beamline focusing on the RFQ beam dynamics.
\end{abstract}

\section{introduction}

Turkish Atomic Energy Authority's (TAEK) Saraykoy Nuclear Research and Education Center (SANAEM) started the Prometheus Project (SPP), which
 aims to construct a proton beamline including a Proof Of Principle (POP) accelerator. The POP machine has the very humble requirements of achieving at least 1.5 MeV proton energy, with an average beam current of at least 1 mA, and also a challenging goal of having the design and construction of the entire machine in Turkey, from its ion source up to last diagnostics station, including its RF power supply and to complete it within three years i.e. by the end of 2015. There are also two secondary goals of this project: 1) To train young accelerator physicists and RF engineers on the job; 2) To involve the local industry in accelerator component construction.

\section{SPP Beamline}

 SPP beamline is approximately 3.5 m long including diagnostics stations. After the extraction of  $H^+$ particles from an ion source at 20 keV, these ions will be send to a solenoid based low energy beam transport (LEBT) line leading  into a Radio Frequency Quadrupole (RFQ) to reach 1.5 MeV which will be followed by a diagnostics station.

\subsection{Ion Source Design}

\begin{figure}[!hbt]
        \centering
        \begin{subfigure}[t]{0.45\textwidth}
     
                \includegraphics[width=77mm]{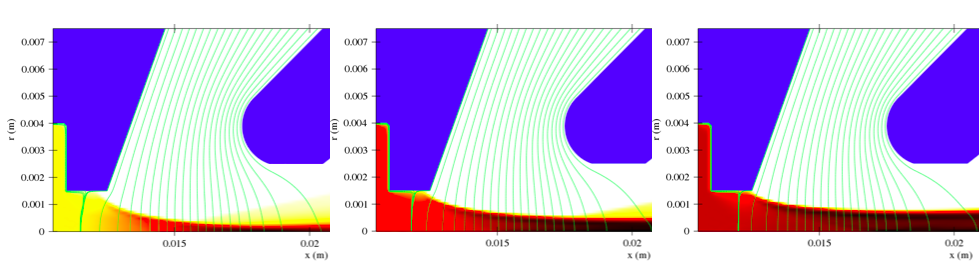}
                \caption{The effects of different current densities on plasma meniscus shapes: Blue regions are electrodes, green lines are equipotential lines, the region in red-yellow represents particle densities.}
                \label{first}
        \end{subfigure} \\
        ~ 
        \begin{subfigure}[b]{0.45\textwidth}
      
                \includegraphics[width=77mm]{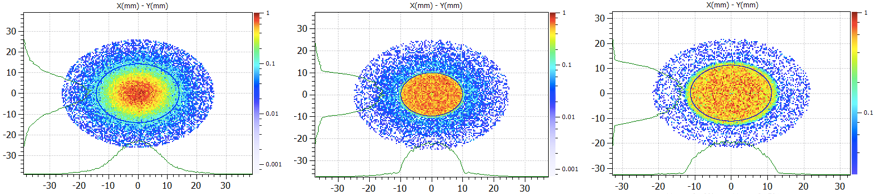}
                \caption{The cross section of output beam of the ion source and the effects of different current densities on beam intensity profiles.}
                \label{second}
        \end{subfigure}
        \caption{Effects of the altering current densities on the ion source properties.}\label{ana}
\end{figure}

A water-cooled multi-cusp  source is designed to supply $H^+$ ions to the RFQ. The output beam energy and emittance of ion source are required to be 20 keV and $<$ 1 $\pi$.mm.mrad respectively, to match with input parameters of  the RFQ. 
The $H^+$ ion source design is made by using IBSimu \cite{ibsimu} computer program which includes a  positive plasma model. The input parameters for  the simulations were the plasma potential, the electron temperature and the current density. Modest parameters were selected for plasma potential and electron temperature. Additionally, Child Langmuir (C-L) Law \cite{Child} was used in the current density calculations. The ion source design was optimized by varying these parameters. As an example, the effects of different current densities on plasma meniscus shapes can be seen in Figure \ref{first}.

Since it is critical to consider the effect of beam intensity distribution and momentum spread on beam emittance growth at the LEBT,  a number of studies were made. If the current density is increased, the beam intensity distribution shape morphs from a Gaussian into a Uniform distribution (Figure \ref{second}). Moreover, investigation shows that the output beam emittance increases when the extraction voltage ranges between 10 kV to 16 kV and it decreases for the range of 16 kV to 20 kV. In addition to these, extraction voltage and beam momentum spread are approximately directly proportional to each other. The final $H^+$ ion source design has therefore 20 keV ion output energy and 16.8 mA beam current with a three electrode extraction system. 
The basic parameters of the ion source are shown in Table \ref{tableion}.

\begin{table}[!hbt]
   \centering
   \caption{Ion Source Parameters}
   \begin{tabular}{lcc}
       \toprule
        Parameter & Value & Unit  \\
       \midrule
        
          Plasma Elect. Aperture Radius & 1.5 & mm    \\
          Extrac. Elect. Aperture Radius & 2.5 & mm    \\
          Plasma-Extraction Elect. Gap & 5.0 & mm     \\
  Plasma Potential & 10.0 & V \\
Electron Temperature & 2.5 & eV \\
Initial Energy of Ions & 5.0 & eV \\
Transverse Temperature of Ions & 0.5 & eV \\
Current Density & 4002.0 & A/$m^2$ \\
Particle Number per Mesh Size & 500 & \# \\
Beam Current & 16.8 & mA \\
Beam Energy & 20.0 & keV \\
Beam Norm. RMS Emittance & 0.0625 & $\pi$.mm.mrad \\
       \bottomrule
   \end{tabular}
   \label{tableion}
\end{table}

\begin{figure}[!htb]
   \centering
   \includegraphics*[width=82mm]{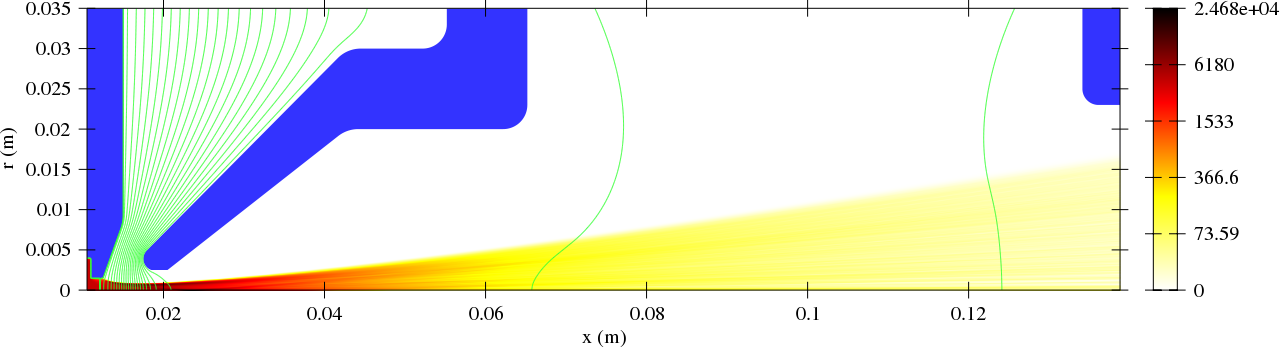}
   \caption{The formation of the beam in extraction system: Electrodes are in blue, equipotential lines are in green.}
   \label{is5}
\end{figure}

 The beam formation inside the three electrode extraction system can be seen in Figure \ref{is5}. The first electrode is kept at 20 kV with plasma chamber while the second and third electrodes are grounded. Thus, 20 kV potential difference occurs at the plasma extraction region. Beside these, the gap between second and third electrodes is reserved for a vacuum pump. 

\begin{figure}[!htb]
   \centering
   \includegraphics*[width=80mm]{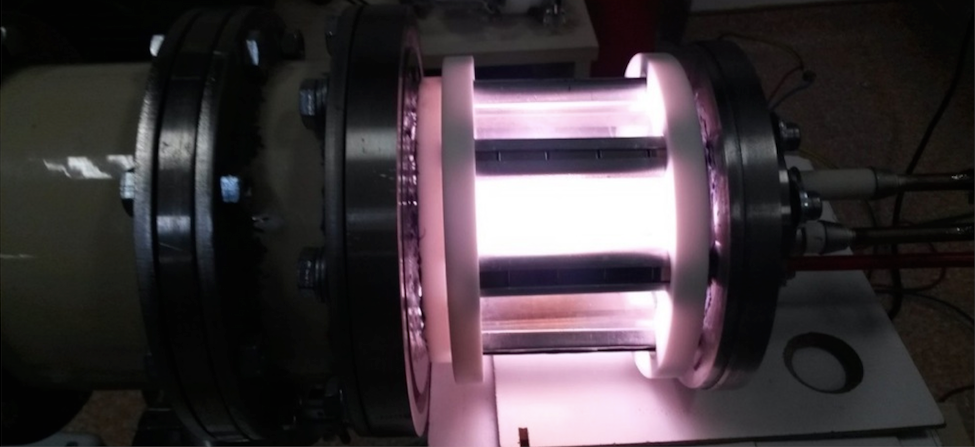}
   \caption{Plasma formation during tests of the ion source.}
   \label{is8}
\end{figure}

A prototype ion source is constructed with cusp magnets and a quartz plasma chamber to visually observe the plasma formation. Plasma tests of the prototype ion source are performed successfully. The ion source has a removable feed-through so that it can be run both at antenna (RF) and filament (DC) modes. 13.56 MHz RF power supply fed  ion source during RF antenna mode tests. Stable plasma is obtained at 230 W forward power. There are 8 row of confinement permanent magnets (3200 G) and their effect on plasma can be seen in Figure \ref{is8}. Plasma wall losses can be slightly eliminated with the help of this magnetic structure.

 Following these tests, water-cooled ion source with metal plasma chamber is under construction. Additionally, plasma and extraction electrode will be built as a removable part for further optimization.

\subsubsection{PFN for Pulsed Beam}
\begin{figure}[!htb]
   \centering
   \includegraphics*[width=80mm]{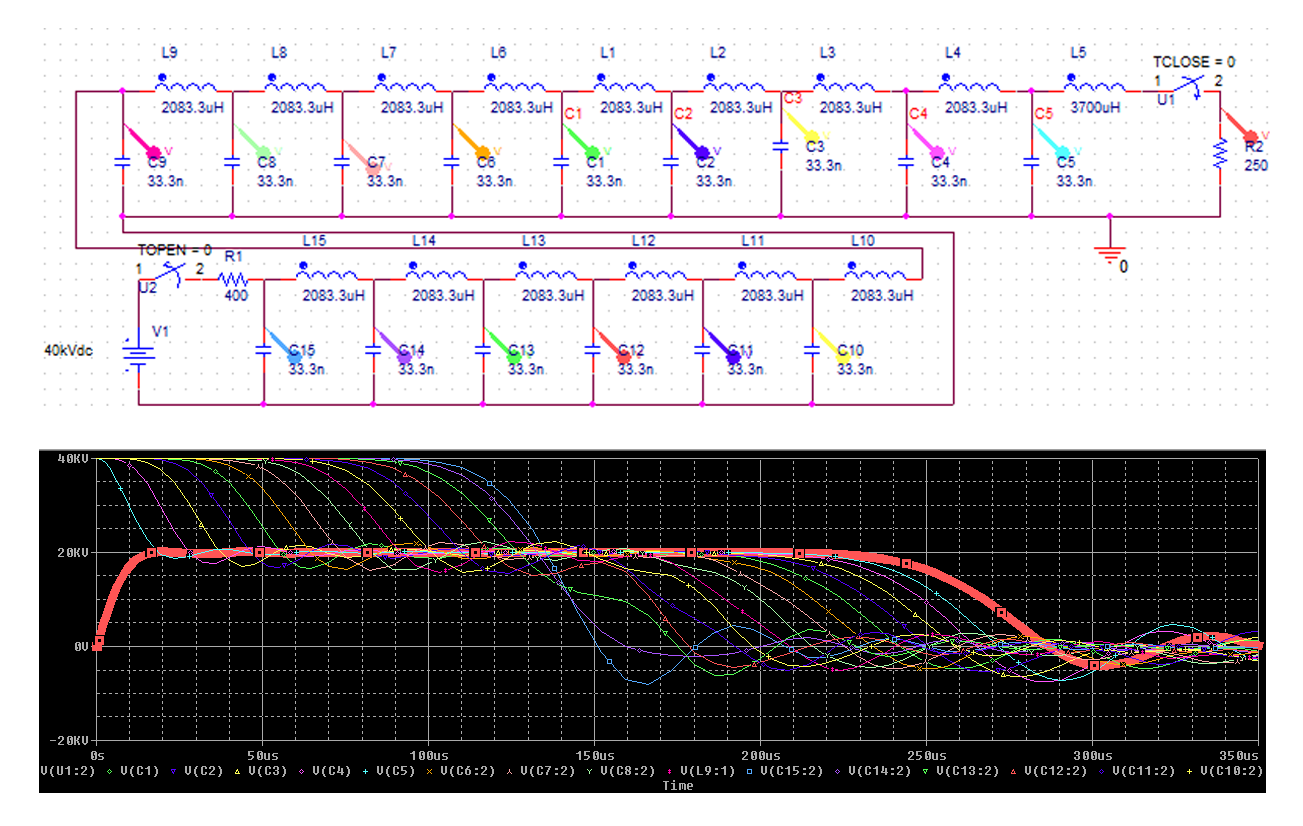}
   \caption{Design of circuit scheme and simulation of the output signal of PFN: Time vs Voltage.}
   \label{is10}
\end{figure}

~\\
As the RFQ will be operated in pulsed mode, so should be the beam injection into it. This will be achieved by a Pulse Forming Network (PFN) which will supply voltage to extraction electrode of the ion source. PFN circuit design and simulation was performed with CADENCE-PSPICE \cite{cadence} software. 
Condensers on the PFN circuit will be charged with 40 kV in order to have the 20 kV extraction voltage. The PFN circuit diagram and the output voltage simulation can be seen in Figure \ref{is10}. In the lower part of this figure, the thin lines represent condenser discharges while the thick red line represents PFN output voltage. As the downstream accelerator components are designed for a specific beam energy, the ion source output should be stable. To fulfill this requirement by making output voltage flat, the inductance of L5 is increased by a factor of 0.437.

\subsection{LEBT Design}
\begin{figure}[!htb]
   \centering
   \includegraphics*[width=60mm]{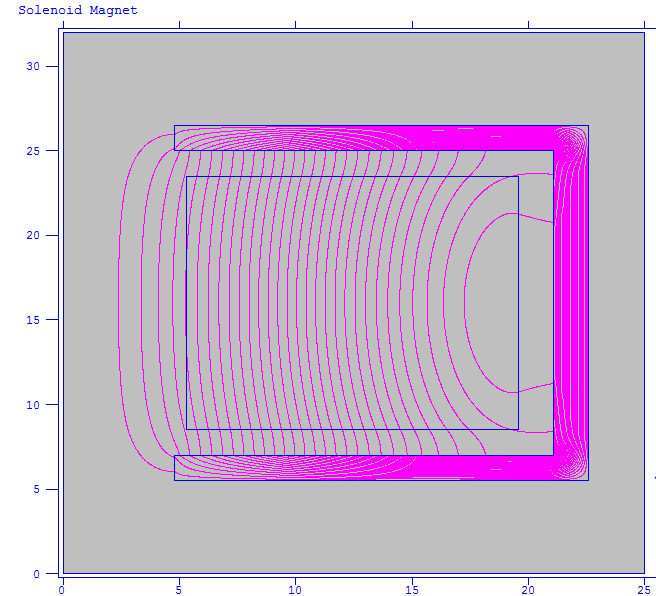}
   \caption{Simulation output of solenoids with POISSON. The pink lines are corresponding to magnetic field lines.}
   \label{solenoid}
\end{figure}

To provide a good match between ion source output and the RFQ input, a LEBT line with two solenoid magnets will be build.
The solenoids  were designed using POISSON \cite{poisson} and FEMM \cite{femm} computer programs. Hard edge solenoid simulations show that the two solenoids should create 2500-2800 G magnetic field for a 50 mm radius beampipe.  Two identical power supplies (100 V, 33 A DC) will feed these two identical solenoids of 
 15 cm  length and 2678 turns of Cu cables that can carry a maximum of 15 A. With these parameters, the solenoids were expected produce a maximum magnetic field of 2740 G (Figure \ref{solenoid}), each. 

 After the completion of the 1.35 kW water-cooled solenoids, the measurements, without cooling, yielded  2600 G magnetic field. The chiller installation is ongoing.

\subsubsection{LEBT Optimization}~\\
The LEBT  was designed with TRAVEL \cite{travel} computer program. Already existing simulation data of ion source output beam and solenoid field maps were used for these simulations. Solenoid field aberrations have been observed because of the low length to aperture ratio. The effect of these aberrations on beam emittance growth has been investigated with the help of multi-particle calculations. More TRAVEL simulations are ongoing with various solenoid strengths and drift lengths. These studies will determine the solenoid currents.

\subsection{RFQ Beam Dynamics Design}

The SPP-RFQ will accelerate 20 keV $H^+$ ions to 1.5 MeV within 1.64 m. RFQ design and beam dynamics simulations are performed by using LIDOS.RFQ.Designer \cite{lidos}, DEMIRCI \cite{demirci} and TOUTATIS \cite{tutos} computer programs. Basic information about SPP RFQ design and comparisons of design softwares can be found elsewhere \cite{Promete1}. Recent machining capability investigations in Turkey, opened up the possibility  for building the entire RFQ from a single piece, instead of from two segments, as thought before. In the two segment case, it is inevitable to have particle losses at the gap region.  The new  design would prevent such losses. Moreover, it has a number of advantages from mechanical and electromagnetic considerations. The simulations for the new design are performed with 50000 macro-particles using a waterbag intensity distribution. The SPP-RFQ parameters are listed in Table \ref{par}.

\begin{table}[!hbt]
   \centering
   \caption{RFQ Main Design Parameters}
   \begin{tabular}{lcc}
       \toprule
        Parameter & Value & Unit  \\
       \midrule
        Resonant Frequency & 352.21 & MHz     \\
        Duty Factor & 2.5 & \%     \\
          Input Energy & 0.02 & MeV    \\
  Beam Current & 1 & mA \\
Input Normalized Emittance & 1 & $\pi$.mm.mrad \\
Inter-vane Voltage & 60 & kV \\
Kilpatrick Limit & 1.5 & - \\
Average Bore Radius & 2.799 & mm \\
t-Rad. of Curv. of V-tip & 2.5 & mm \\
Output Energy & 1.5 & MeV    \\
Output Norm. t-Emit. (x-y) & 1.56-1.50 & $\pi$.mm.mrad \\
Output Norm. l-Emit. & 0.4 & MeV*deg \\ 
Transmission (Acc) & 96.2 & \% \\ 
Total Length & 164.6 & cm \\ 
       \bottomrule
   \end{tabular}
   \label{par}  
\end{table}

The effects of the beam intensity distributions (Uniform, Gaussian, Waterbag) on transmission were also investigated to determine the accelerated particle transmission percentages as 95.8\%, 71.0\% and 96.2\% for uniform, Gaussian and waterbag intensities, respectively. It is easily seen that transmission is low for Gaussian intensity due to the long tails of Gaussian distribution. Although maximum electric field in SPP-RFQ is 21 MV/m (corresponding to 60 kV), the $H+$ ions can reach 1.5 MeV with 96.2\% transmission within 1.64 m.

\subsubsection{Tolerances}~\\
Influence of the input parameters of RFQ such as beam energy, current, emittance and RFQ inter-vane voltage on transmission are investigated by using various LIDOS simulations. Tolerances of the  RFQ input parameters will be determined and effect on transmission will be predicted with the help of these studies. An 85\% transmission of  accelerated particles could be considered acceptable for a POP machine. Various other tolerances are listed in Table \ref{tole}.

\begin{table}[!hbt]
   \centering
   \caption{Tolerances}
   \begin{tabular}{lccc}
       \toprule
        Parameter & Value & Tolerance & Unit  \\
       \midrule
        Beam Energy & 20 & -1.75 / +2.5 & keV   \\
        Beam Current & 1 & $<$ 19.5 & mA     \\
        Beam Emittance & 1 & $<$ 1.7 & $\pi$.mm.mrad    \\
  Inter-vane Voltage & 60 & $>$ 56 & kV \\
       \bottomrule
   \end{tabular}
   \label{tole}
\end{table}

\section{Conclusion}

This paper  summarizes the status of the SPP beamline, including its design, simulations and the results of the prototype tests. 
The ion source and the LEBT are currently being built. 
There is a new RFQ design based on a single piece manufacturing under consideration. If the project timeline is kept, the first MeV range protons are
expected by the end of 2015.
 
\section{acknowledgment}
This project is funded by TAEK with a project code A1.H4.P1.03. We would like to thank Taneli Kalvas and Ece Asilar for many helpful discussions.


\end{document}